\documentclass[10pt,a4]{article}
\usepackage[centertags]{amsmath}
\usepackage{amssymb}
\usepackage{amsfonts}
\usepackage{graphicx,color}
\usepackage{epsfig}
\usepackage{amsxtra}
\usepackage{amstext}
\usepackage{latexsym}
\usepackage{epstopdf}
\usepackage{makeidx}
\def\be{\begin{equation}}
\def\ee{\end{equation}}
\def\bea{\begin{eqnarray}}
\def\eea{\end{eqnarray}}

\begin{document}
\begin{center}
{\Large {\bf Centrality of L\"owdin orthogonalizations}}\\
\vskip 1cm
{\bf Annavarapu Ramesh Naidu$^\dagger$}\\
\vskip .1cm
Department of Physics\\ School of Physical, Chemical and Applied Sciences\\
Pondicherry University, Puducherry 605 014, India.\\
\vskip1cm
$^\dagger$ rameshnaidu.phy@pondiuni.edu.in
\vskip 3cm
\textbf{Abstract}\\
\vskip .2cm
\end{center}

The different orthogonal relationships that exists in the L\"{o}wdin orthogonalizations is presented. Other orthogonalization techniques such as polar decomposition (PD), principal component analysis (PCA) and reduced singular value decomposition (SVD) can be derived from L\"{o}wdin methods. It is analytically shown that the polar decomposition is presented in the symmetric orthogonalization; principal component analysis and singular value decomposition are in the canonical orthogonalization. The canonical orthogonalization can be brought in into the form of reduced SVD or vice-versa. The analytic relation between symmetric and canonical orthogonalization methods is established. The inter-relationship between symmetric orthogonalization and singular value decomposition is presented. 

\newpage
\noindent {\bf Introduction}\\

Orthogonalization methods come in two categories, viz., \emph{sequential} and \emph{democratic}. The Gram-Schmidt method \cite{1} takes the linearly independent set of vectors one-by-one and gives an orthonormal set. The \emph{democratic} orthogonalization methods due to L\"owdin namely the \emph{symmetric} and \emph{canonical} orthogonalizations handle all the given vectors simultaneously and treat them on equal footing. L\"{o}wdin orthogonalization methods \cite{2, 4} were discovered for the purpose of orthogonalizing hybrid electron orbits in quantum chemistry. A few other orthogonalization methods have been developed independently to deal with specific problems in computer science, mathematics, statistics and biology etc.
The centrality of L\"owdin orthogonalization schemes to other orthogonalization techniques such as polar decomposition, principal component analysis and singular value decomposition is presented. The derivation and interesting geometric properties of the two procedures are described in the references \cite{2, 5, 6}. Applications of L\"owdin methods to a variety of interdisciplinary problems have been explored in the last decade. They have applications in cognitive phenomena \cite{7, 8}, data reduction \cite{9} and in the generation of new polynomials \cite{10}.\\

\noindent {\bf L\"owdin Orthogonalizations}\\

Let ${\bf V}\equiv \{\vec{v}_1,\vec{v}_2,\cdots,\vec{v}_N\}$ be a a set of linearly independent vectors in a $N$-dimensional space which can in general be a complex vector space.
A general non-singular linear transformation ${\bf A}$ can transform the basis ${\bf V}$ to a new basis ${\bf Z}$:
\begin{equation}
{\bf Z}={\bf V}{\bf A}.
\end{equation}
The set ${\bf Z}(\equiv \{\vec{z}_\kappa\})$ will be orthonormal if
\begin{equation}
<{\bf Z}|{\bf Z}>=<{\bf VA}|{\bf VA}>={\bf A}^{\dagger}<{\bf V}|
{\bf V}>{\bf A}={\bf A}^{\dagger}{\bf M}{\bf A}={\bf I},
\end{equation}
where ${\bf M}$ is a Hermitian metric matrix of the given basis {\bf V}.

A general solution to the orthogonalization problem is obtained using the substitution
\begin{equation}
{\bf A}={\bf M^{-1/2}}{\bf B},
\end{equation}
where ${\bf B}$ is an arbitrary unitary matrix.

The specific choice ${\bf B}={\bf I}$ gives the {\it symmetric orthogonalization},
\begin{equation}
{\bf Z}\equiv{\bf \Phi}={\bf V}{\bf M^{-1/2}}, \label{eq:4}
\end{equation}
while ${\bf B}={\bf U}$, where ${\bf U}$ diagonalizes ${\bf M}$ gives the {\it canonical orthogonalization},
\begin{equation}
{\bf Z}\equiv{\bf \Lambda}= {\bf V{\bf U}{\bf d}^{-1/2}}.
\end{equation}

These two orthogonalized basis sets have novel geometric properties \cite{5} when we consider the Schweinler - Wigner matrix \cite{6} in terms of the sum of squared projections.\\

\noindent{\bf Polar Decomposition}\\

The polar decomposition \cite{11} of a matrix $\mathbf{V} \in \mathbb{R}^{n \times N}$, where $\mathbb{R}$ is a set of real numbers, can be obtained from the L\"{o}wdin's symmetric orthogonalization. The symmetric orthonormal basis of a matrix $\mathbf{V}$ is expressed as
\be
\mathbf{\Phi = V M^{-1/2}}, \label{eq:7}
\ee
where $\mathbf{\Phi} \in \mathbb{R}^{n \times m}$ and $\mathbf{M} \in \mathbb{R}^{m \times m}$.
Now multiplying both sides of eqn. (\ref{eq:7}) from right by $\mathbf{{M}^{1/2}}$, we get
\bea
\text{or}, \mathbf{V} &=& \mathbf{\Phi M^{1/2}}. \label{eq:5.2}
\eea
This is called as polar decomposition of the matrix $\mathbf{V}$.\\

\noindent{\bf Principal Component Analysis}\\

Let $\mathbf{V} \in \mathbb{R}^{N \times n}$, then the sum of squares and cross-prodcuts (SSCP) matrix $\mathbf{S}$ can be constructed using $\mathbf{V V^{T}}$. The SSCP matrix is the covariance matrix without subtracting the mean. The diagonal values are sums of squares and the off-diagonal values are sums of cross products. The eigenvalues  and eigenvectors of the SSCP matrix $\mathbf{S}$ are constructed. We have found that the eigenvalues are the same as the eigenvalues of the Gram matrix constructed using $\mathbf{V^{T} V}$ in the case of L\"owdin orthogonalizations. The eigenvectors in the two cases are different. However, the eigenvectors of the SSCP matrix $\mathbf{S}$, called as the principal components \cite{12} of $\mathbf{V}$, are the same as those obtained using the canonical orthogonalized set $\mathbf{\Lambda = V U d^{-1/2}}$. The eigenvectors of $\mathbf{S}$ are ordered so that the first two principal components retain most of the variance present in the original set of vectors. We have computed the sum of the projection-squares \cite{5} of the given vectors onto principal components \cite{5, 13}. This gives the eigenvalues of $\mathbf{S}$ and is the same as the sum of projection-squares of the original vectors on the canonical orthonormal vectors, i.e., the eigenvalues of $\mathbf{M}$.

For $N$ vectors in $n$ dimensions, we find that the principal components obtained through the principal component analysis of a square matrix $\mathbf{V}$ are equal to the orthonormal vectors obtained through the L\"owdin's canonical orthogonalization. Hence in the case of square matrices, the principal component analysis of the pure SSCP matrix is equivalent to the canonical orthogonalization.\\
\\

\noindent{\bf Singular Value Decomposition}\\

The singular value decomposition \cite{14, 15} of {a} non-singular matrix $\mathbf{V}$ can be obtained from the L\"owdin's canonical orthogonalization. The canonical orthogonalization of a matrix $\mathbf{V}$ can be written as
\bea
\mathbf{\Lambda} &=& \mathbf{V U d^{-1/2}}. \label{eq:9}
\eea
Multiplying both sides of equation (\ref{eq:9}) from its right with $\mathbf{d^{1/2}}$, we have
\bea
\mathbf{\Lambda d^{1/2}} &=& \mathbf{V U~ I = V U}. \label{eq:10}
\eea
Now multiplying equation (\ref{eq:10}) on both sides from right with $\mathbf{U^{\dagger}}$, we get
\bea
\mathbf{V} &=& \mathbf{\Lambda d^{1/2} U^{\dagger}} \label{eq:11}
\eea
This is the singular value decomposition of matrix $\mathbf{V}$ and is called as reduced singular value decomposition form of the canonical orthogonalization.\\

\noindent{\bf Analytic Relations between Symmetric and Canonical Orthogonalizations}\\

We can analytically obtain the relationship between symmetric and canonical orthogonalizations. If one of them is obtained from the Hermitian metric matrix, say canonical or symmetric, then the other can be obtained from the following fundamental relations.

The symmetric orthonormal basis is given by
\bea
\mathbf{\Phi = V M^{-1/2}}.
\eea
From the equation $\mathbf{Z = V A = V M^{-1/2} B}$ and for $\mathbf{B = U}$, We have
\bea
\mathbf{\Lambda} &=& \mathbf{V M^{-1/2} U}\\
\text{Or,}~\mathbf{\Lambda} &=& \mathbf{\Phi U}. \label{eq:15}
\eea
This analytic relation is useful to construct the canonical orthonormal basis directly using symmetric orthonormal basis and the eigenvectors of the Hermitian metric matrix.

The symmetric orthonormal basis can be obtained from the canonical orthonormal basis by multiplying both sides of equation (\ref{eq:15}) from its right with $\mathbf{U^{\dagger}}$ as shown below.
\bea
\mathbf{\Lambda U^{\dagger}} &=& \mathbf{\Phi ~{U U^{\dagger}}} \\
\mathbf{\Lambda U^{\dagger}} &=& \mathbf{\Phi I = \Phi}. \\
\text{Or,}~ \mathbf{\Phi} &=& \mathbf{\Lambda U^{\dagger}}.
\eea

\noindent{\bf Symmetric Orthogonalization in SVD}\\

We can analytically obtain the symmetric orthogonalization from the singular value decomposition. Let $\mathbf{V \in \mathbb{R}}^{n \times m}$ with $n \ge m$ be a non-singular matrix. Then the singular value decomposition of $\mathbf{V}$ can be written as
\bea
\mathbf{V} &=& \left(
  \begin{array}{cccc}
    w_{11} & w_{12} & \cdots & w_{1m} \\
   w_{21} & w_{22} & \cdots & w_{2m}\\
    \vdots & \vdots & \ddots & \vdots  \\
    w_{n1} & w_{n2} & \cdots & w_{nm} \\
  \end{array}
\right) \left(
  \begin{array}{cccc}
    \sigma_{1} & 0 & \cdots & 0 \\
   0 & \sigma_{2} & 0 & 0\\
    \vdots & \vdots & \ddots & \vdots  \\
    0 & 0 & \cdots & \sigma_{m} \\
  \end{array}
\right) \left(
  \begin{array}{cccc}
    u_{11} & u_{12} & \cdots & u_{1m} \\
   u_{21} & w_{22} & \cdots & u_{2m}\\
    \vdots & \vdots & \ddots & \vdots  \\
    u_{m1} & u_{m2} & \cdots & u_{mm} \\
  \end{array}
\right) \nonumber\\
\nonumber\\
 &\equiv& \mathbf{W~\Sigma~U^{\dagger},~~\text{where}~(\Sigma = d^{1/2})}, \nonumber
\eea
where $\mathbf{W \in \mathbb{R}}^{n \times m}$ and $\mathbf{U \in \mathbb{R}}^{m \times m}$ are matrices with their columns as orthonormal vectors. The columns $\mathbf{W}_{j},~j=1, 2,  \cdots, m$ of $\mathbf{W}$ are called {\em left singular vectors} of $\mathbf{V}$ and the columns of $\mathbf{U}_{j},~j=1, 2,  \cdots, m$ of $\mathbf{U}$  (or rows of $\mathbf{U^{\dagger}}$)are called {\em right singular vectors} of $\mathbf{V}$. And $\mathbf{\Sigma \in \mathbb{R}}^{m \times m}$ is square and diagonal matrix with $\mathbf{\sigma}_{i}$'s as the {\em singular values} of $V$. By convention, the singular values are arranged in a descending order as $\mathbf{\sigma}_{1} \ge \mathbf{\sigma}_{2} \ge \cdots \ge \mathbf{\sigma}_{m} \ge 0$. This form of singular value decomposition is known as {\em reduced singular value decomposition}. This reduced SVD can give the symmetric orthogonalization. Using the analytic relationship between the symmetric and canonical orthogonalizations, the symmetric orthogonalization of the matrix $\mathbb{V}$ can be written from the reduced SVD as follows,
\bea
\mathbf{\Phi} &=& \mathbf{W U^{\dagger}}, \\
\mathbf{\Phi}_{ij} &=& \mathbf{W}_{i1} \left(\mathbf{U}^{\dagger} \right)_{1j}+ \mathbf{W}_{i2} \left(\mathbf{U}^{\dagger} \right)_{2j}+ \cdots + \mathbf{W}_{im} \left(\mathbf{U}^{\dagger} \right)_{mj} \\
&=& \mathbf{W}_{i1}\left(\mathbf{U}\right)_{j1}+\mathbf{W}_{i2}\left(\mathbf{U} \right)_{j2}+ \cdots + \mathbf{W}_{im}\left(\mathbf{U} \right)_{jm},
\eea
where $\mathbf{W}$ is {the} same as the canonical orthonormal basis $\mathbf{\Lambda}$. The symmetric orthonormal basis $\mathbf{\Phi}$ is {\em unique} since it takes the linearly independent set of columns of $\mathbf{V}$ as input and gives the orthonormal columns of $\mathbf{W U^{\dagger}}$ as output.
\\

\noindent{\bf Discussion}\\

The {\em democratic} orthogonalization procedures considers the entire lot of vectors in one go. We concentrated mainly on the democratic type because their applications have not been explored much except in the domain of quantum chemistry. We have established that the canonical orthogonalization was in fact invented independently by several people from time to time with different names such as principal component analysis and singular value decomposition. We have established their equivalence. The connection between symmetric and canonical orthogonalizations have also been established. Analytic relation between symmetric and canonical orthogonalizations methods is presented. The inter-relationship between symmetric and SVD is also shown.\\

\noindent{\it Acknowledgements:} This work is financially supported from the grant of University Grants Commission (UGC), New Delhi, India. The author would like to thank the UGC for the financial support.\\

\end{document}